# PET image reconstruction with system matrix containing point spread function derived from single photon incidence response*

Fan Xin(樊馨)[a)c)], Wang Hai-Peng(王海鹏)[a)c)], Yun Ming-Kai(贠明凯)[a)b)], Sun Xiao-Li(孙校丽)[a)c)], Cao Xue-Xiang(曹学香)[a)b)], Liu Shuang-Quan(刘双全)[a)b)], Chai Pei(柴培)[a)b)], Li Dao-Wu(李道武)[a)b)], Liu Bao-Dong(刘宝东) [a)b)],Wang Lu(王璐) [a)c)] and Wei Long(魏龙)[a)b)†]

[a)] Key Laboratory of Nuclear Analytical Techniques, Institute of High Energy Physics, Chinese Academy of Sciences, Beijing 100049, China

[b)] Beijing Engineering Research Center of Radiographic Techniques and Equipment, Beijing 100049, China

[c)] University of Chinese Academy of Sciences, Beijing 100049, China

## Abstract

In positron emission tomography (PET) imaging, statistical iterative reconstruction (IR) techniques appear particularly promising since they can provide accurate system model. The system model matrix which describes the relationship between image space and projection space is important to the image quality. It contains some factors such as geometrical component and blurring component. The blurring component is usually described by point spread function (PSF). A PSF matrix derived from the single photon incidence response function is studied. And then an IR method based on the system matrix containing the PSF is developed. More specifically, the gamma photon incidence on a crystal array is simulated by Monte Carlo (MC) simulation, and then the single photon incidence response functions are calculated. Subsequently, the single photon incidence response functions is used to compute the coincidence blurring factor according to the physical process of PET coincidence detection. Through weighting the ordinary system matrix response by the coincidence blurring factors, the IR system matrix containing PSF is finally established. Using this system matrix, the image is

*Project supported by National Natural Science Foundation of China (Grant No Y4811H805C and 81101175).

† Corresponding author. E-mail: weil@ihep.ac.cn

reconstructed by ordered subset expectation maximization (OSEM) algorithm. The experimental results show that the proposed system matrix can obviously improve the image radial resolution, contrast and noise property. Furthermore, the simulated single gamma-ray incidence response function only depends on the crystal configuration, so the method could be extended to any PET scanners with the same detector crystal configuration.

Key words: PSF, single photon incidence, system matrix, PET

**PACS**: 87.57.nf, 87.57.uk, 87.57.C-

## 1. Introduction

Positron emission tomography (PET) is a nuclear medical imaging technique and provides important information for disease diagnosis, therapeutic effect assessment and new drug development.[1] The PET system detects pairs of back to back gamma photons emitted indirectly from a positron-emitting radionuclide, which is injected into the living body on a biologically active tracer. The image of tracer concentration within the living body can be acquired by image reconstruction methods such as analytic reconstruction [2] and statistical iterative reconstruction (IR).[3] The quality of PET imaging is vital for disease diagnosis and evaluation of treatment. It includes image resolution, contrast, noise property and so on.[4] The image resolution which is crucial for the diagnosis of early stage tumor mainly depends on several factors such as the size of detector, the photon non-colinearity, the positron range and inter-crystal penetration.[5,6,7,8] The size of detector may not be changed for an existing system. Among the other physical and geometric factors, crystal penetration will lead to depth-of-interaction (DOI) blurring. The image spatial resolution degradation and positional error turn more serious as the DOI blurring increasing.[9] These physical and geometric factors can be accurately modeled by the system matrix in IR reconstruction.[8]

Traditionally, the system matrix can be divided into some component such as geometrical component, blurring component and so on. Point spread function (PSF) is generally used to describe the blurring component. PSF can be modeled by the analytic methods, [10-12] the Monte Carlo (MC) simulation methods[13-15] and the experimental methods.[16-22] It's a huge work to get the spatially variant PSFs[23] of all the voxels in the experiment methods,[19] while purely analytic methods is less accurate than the other methods.[19][20] Several studies proposed some methods which firstly obtained a few specific PSFs by the experimental measurements or MC simulations and then used specific models ( for example Gaussian function model[19] or iterative algorithm[22]) to estimate the PSF of all voxels based on the system symmetry.[19][20][22] Thus, the experimental time is reduced dramatically. However, it's improper for an accurate system model to only model the radial blurring but ignore the azimuthal blurring [19][20] as in most of these methods. Moreover, it is tedious work that more than one experiment or simulation needs to be implemented for different geometrical structures of PET scanners in these methods.[19][20]

In this paper, we propose a new method which calculates the PSF information based on the single gamma photon incidence response function. PET imaging theory is introduced in section 2. The new method is introduced in section 3. The improved results and corresponding analysis is displayed in section 4. Finally, in section 5, some related problem is discussed.

## 2. PET imaging theory

In IR methods, system matrix describes the relationship between projection and image space. This relationship can be expressed as

$$p_j = \sum_i a_{ij} f_i \, , \tag{1}$$

where $p_j$ is the true value of the projection data for a line of response (LOR) decided by the detector pair $j$. $f_i$ is the value of image at voxel $i$. $a_{ij}$ is the probability of detecting a coincidence event originating from voxel $i$ at detector pair $j$.[19] We define $A$ as the matrix of $a_{ij}$.. So $A$ represents the system matrix which can be divided into several factored matrices.[19] It can be expressed as

$$A = A_{sens} A_{atten} A_{blur} A_{geom} A_{positron} \quad . \tag{2}$$

The positron range factor $A_{positron}$ is relatively smaller and can be ignored for $^{18}F$.[24] The attenuation factor $A_{atten}$ can be provided by an extra CT scan[25] and the detector sensitivity factor $A_{sens}$ can be acquired by measuring a uniform cylindrical source.[26] The remaining factors are geometrical factor $A_{geom}$ and the blurring factor $A_{blur}$.[27] The geometrical factor can be accurately estimated by some analytical methods such as line integral model[28], tube model[29] and solid angle model.[30][19] However, owing to the complicated response, it is always difficult to acquire accurate blurring factor. It should include physical effects such as crystal penetration and photon non-colinearity[5,6,7,8] which would result in the degradation of reconstructed image quality as mentioned before.

# 3. PSF modeling method

## 3.1. Monte Carlo simulation for single photon incidence response function

Generally, modern PET scanner uses uniform detector blocks and has a polygonal shape. The block consists of a crystal array to which a number (usually four) of photomultiplier tubes are attached. Fig.1 shows the PET structure and block structure.

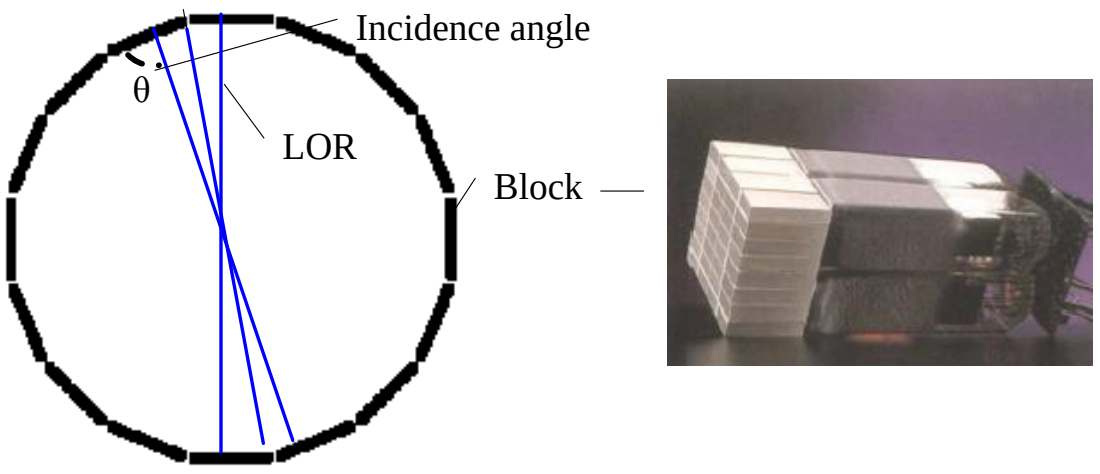

Fig.1. (color online) PET structure and its block

In PET system, blurring effect mainly refers to DOI blurring. And DOI blurring results from crystal penetration which is caused by non-normal incidence of gamma-ray.[9] The bigger the angle of incidence is, the more serious DOI blurring is.[31] Fig.2 displays three different incidence patterns of gamma-ray to the detector array. In Fig2 (a), there are few crystals being penetrated for normal incidence. In Fig.2 (b) and (c), the gamma-ray may penetrate a few adjacent crystals for non-normal incidence. The response of gamma-ray penetrating crystals is mainly decided by incidence angle for the same crystal configuration.[31] In Fig.2 (b), two crystals are penetrated while in Fig.2 (c) whose incidence angle is larger, three crystals are penetrated. Theoretically, because blocks in PET are uniform, the responses of blocks are the same. So, we only need to study the responses of all incidence angles for one block.

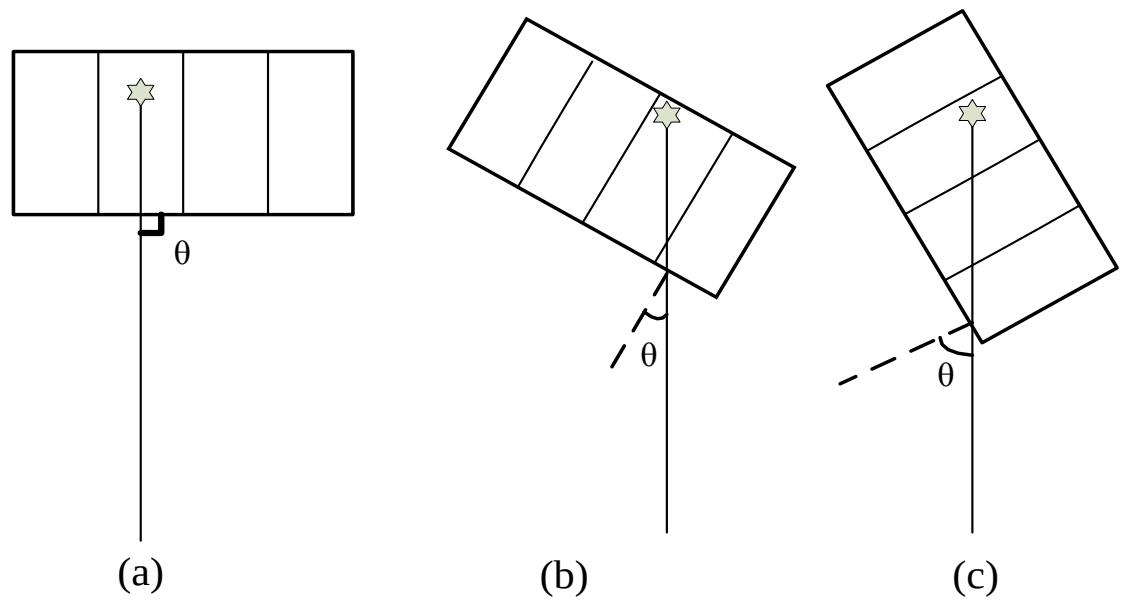

Fig.2. Three different incidence patterns of gamma-ray. (a) vertical incidence. (b) oblique incidence. (c) larger angle of incidence compared to (b)

The incidence angle range in the simulated experiment is set according to that of the existing device. Here, we defined the complement angle of the ordinary incidence angle as the incidence angle for the convenience of calculation (for example θ of Fig.3). We calculated the incidence angle range for system geometry of sixty-four polygon with 11×11 LYSO crystals in each block. The crystal size was 3.5 mm×3.5 mm×15 mm. The gap between every two blocks was 4 mm. Incidence angle range of single photon for the system is 42.7 °- 90.0 °with 353 bins of every angle for transverse plane. So we can get the response of incidence angles from 30 °to 90 °to satisfy incidence angles of this scanning system.

We simulated this single photon incidence response mentioned above by Geant4 Application for Emission Tomography (GATE) software based on MC methods.[32] As shown in Fig.3, in the simulation, 15×1 LYSO crystal array was set behind a lead collimator. The crystal array was rotated to produce different angles and the lead collimator assured the direction of single incidence. We took 5° for incidence angle-step from 30° to 90° for the first attempt. The response was obtained based on probabilistic method. As shown in Fig. 4, assuming N events have been recorded totally, there are n1, n2 and n3 events being recorded in crystal 1, crystal 2 and crystal 3 respectively, so the corresponding probability of response are n1/N, n2/N and n3/N.

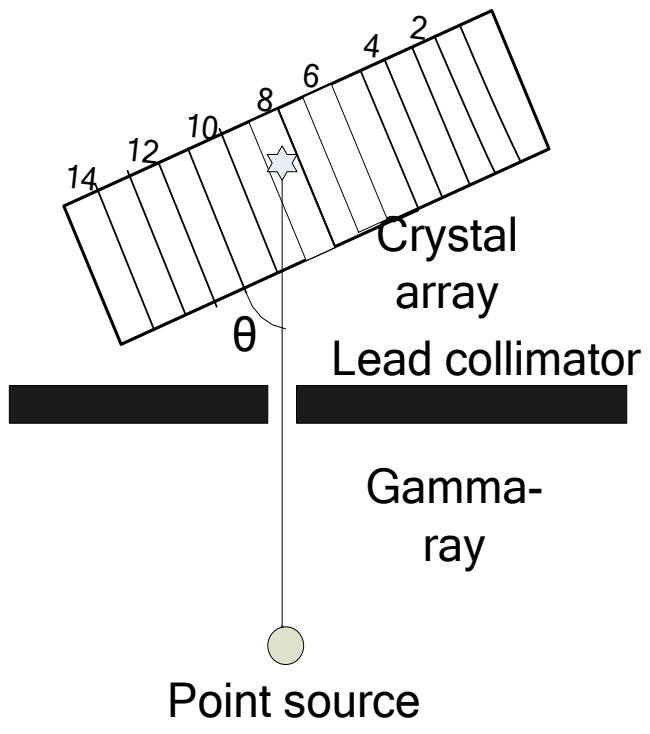

Fig.3. The setting of MC simulation of single photon incidence

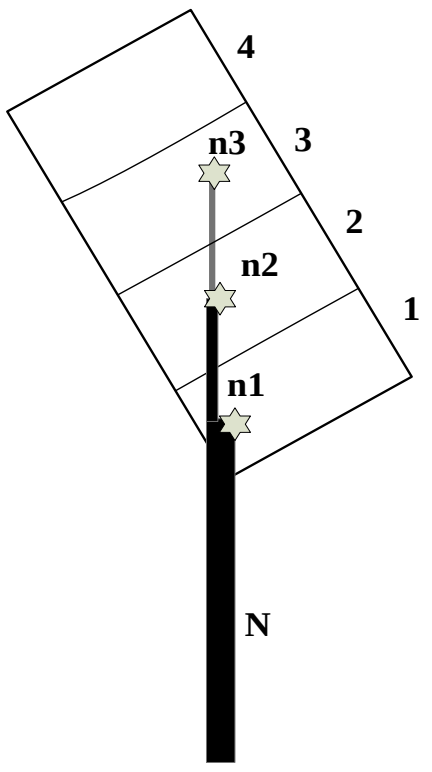

Fig.4. Probabilistic method of getting the response of single photon incidence

Fig.5 displays the responses of crystal array with two different simulated incidence angles. Fig.5 (a) and (c) show the original response. To save the reconstruction time, the probability of each crystal is accumulated along the detected order. If the sum ⩾ 80%, the probabilities of the rest crystals are set to zeros and the accounted probabilities are normalized with the sum. Fig.5 (b) and (d) show the normalized response. We can see that gamma-ray of 60° incidence angle has penetrated more crystals than gamma-ray of 90° angle (vertical incidence), which is the same as a previous study.[31].

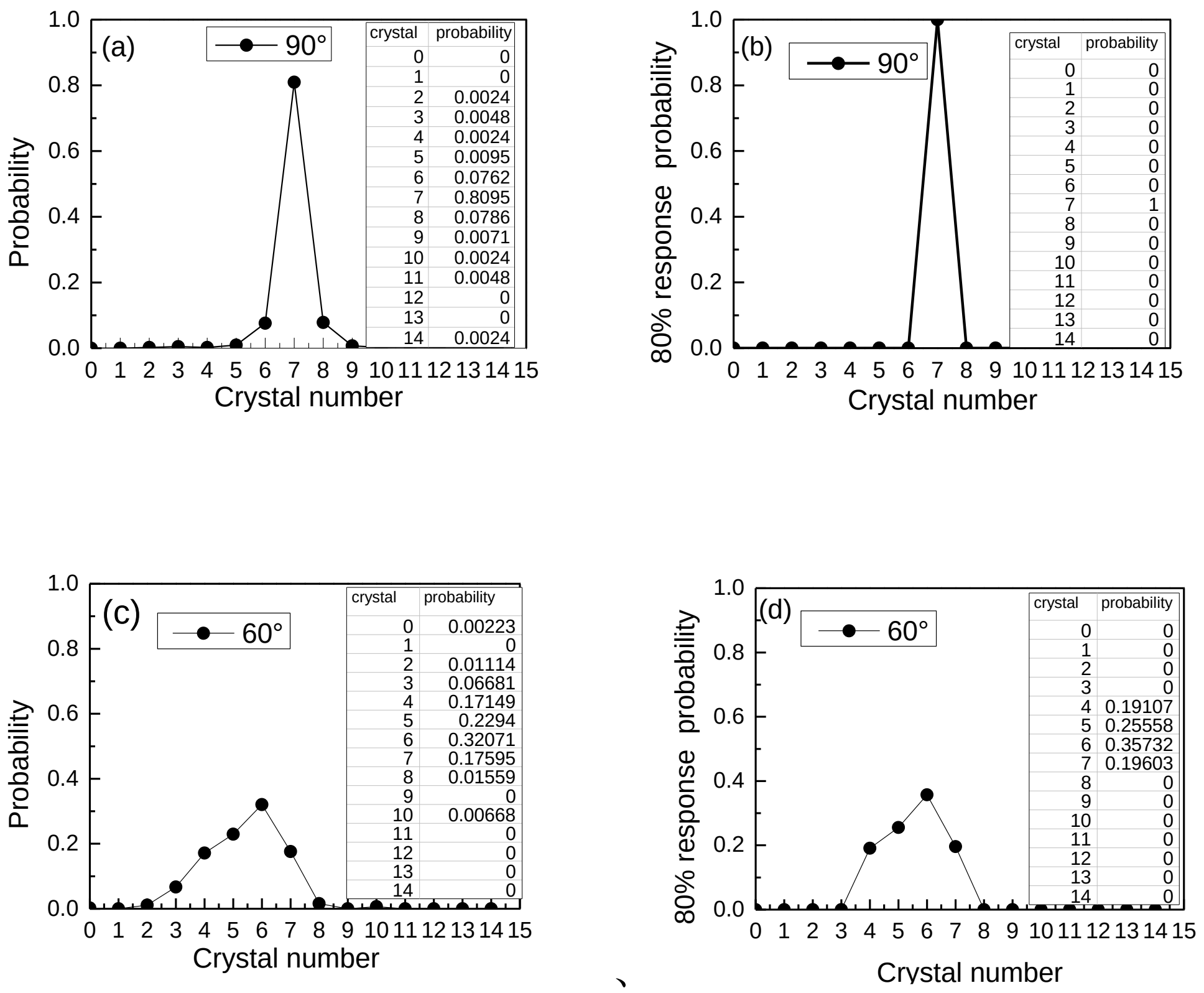

Fig.5. Single photon incidence response function simulated by MC methods. (a) and (c) are the original response of 90 ° and 60 ° incidence angle respectively. (b) and (d) are the normalized responses corresponding to (a).and (c). The inset shows the corresponding response table.

## 3.2. Coincidence for PSF-based system matrix

In PET system, raw data is acquired from coincidence events of crystal pairs. When coincidence response is produced from a pair of crystals, single event responses of two crystals are generated simultaneously, so that response signals of penetrated crystals in two sides would be recorded as a coincidence event for every two crystals within the coincidence timing window. [31]. As shown in Fig.6, we coincided two back to back single gamma-ray photon responses using the single photon incidence response function in section 3.1 according to this physical process. As in Fig.6, solid line represents incidence coincidence LOR, and incidence angles of corresponding crystal *b* and *g* are $\theta_1$ and $\theta_2$ respectively. On the side of $\theta_1$, penetrated crystals are *b* and *c*, while on the other side penetrated crystals are *g*, *h* and *i*. Thus, blurrings of coincidence LOR are decided by responses in (*b*, *g*), (*b*, *h*), (*b*, *i*), (*c*, *g*), (*c*, *h*) and (*c*, *i*) crystal pairs.

And we take the product of probability of the two sides penetrated crystals in the single photon response function as the corresponding blurring response probability.[31] Table 1 shows the probabilities of $\theta_1$ and $\theta_2$ in single photon response function. Crystal *b* and *g* is the incident crystal. So for incidence of $\theta_1$, response probabilities of penetrated crystal *b* and *c* are $p_{11}$ and $p_{12}$ respectively according to table 1. Similarly, the response probabilities of crystal *g*, *h* and *i* are $p_{21}$, $p_{22}$, $p_{23}$ respectively. Table 2 shows the coincident results of LOR blurring response. Obviously, LOR blurring response contains both radial and azimuthal blurring in transverse plane.[23]

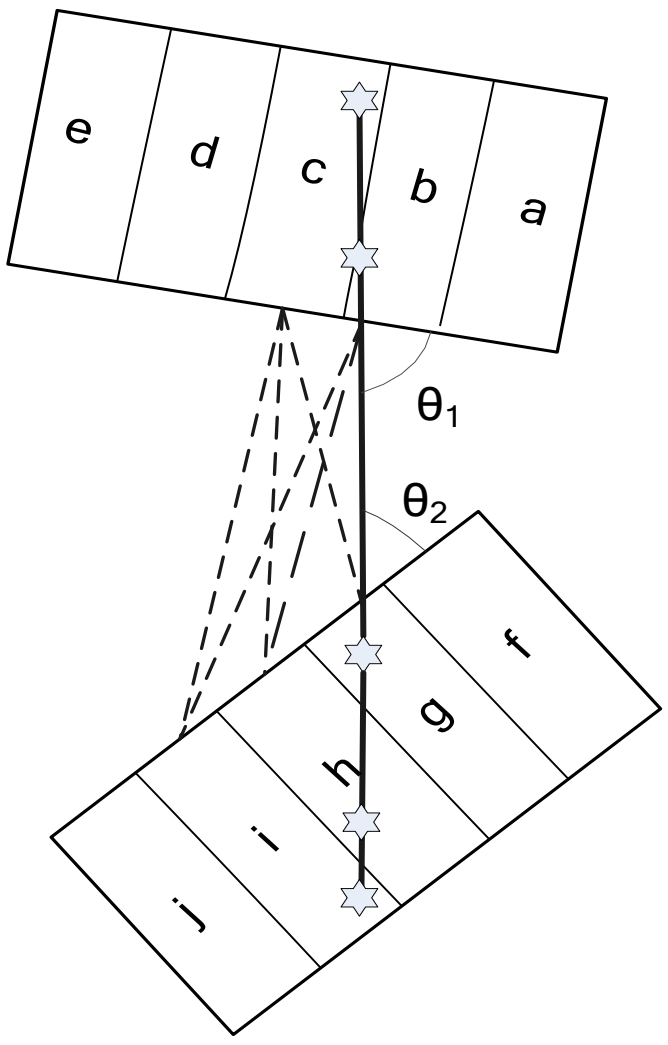

Fig.6. Coincidence process. Solid line is the incident coincidence LOR and dash lines are LORs which cause the blurring.

**Table 1.** Single photon response probabilities of incidence angle θ1 and θ2.

| Incident angle | Crystal number | | | | |
|---|---|---|---|---|---|
| | *b* | *c* | *g* | *h* | *i* |
| $\theta_1$ | $p_{11}$ | $p_{12}$ | / | / | / |
| $\theta_2$ | / | / | $p_{21}$ | $p_{22}$ | $p_{23}$ |

**Table 2.** The coincident result of LOR blurring response.

| Coincident event | Response probability |
|---|---|
| Event (*b*,*g*) | $p_{11} \times p_{21}$ |
| Event (*b*,*h*) | $p_{11} \times p_{22}$ |
| Event (*b*,*i*) | $p_{11} \times p_{23}$ |
| Event (*c*,*g*) | $p_{12} \times p_{21}$ |
| Event (*c*,*h*) | $p_{12} \times p_{22}$ |
| Event (*c*,*i*) | $p_{12} \times p_{23}$ |

To simplify the calculation, we can only calculate a certain number of LOR blurring responses according to the symmetry properties of PET system geometry.[22]

Fig.7 shows coincident result of sinogram (an organizational form of LOR) blurring of two different radial positions in angle of 0 °.We also use the scanner system which has sixty-four polygon geometry as mentioned above. We can see that these blurring of both two positions include radial blurring and azimuthal blurring. (a) and (b) represent the response of the furthest bin from the center of the FOV. The bin of the biggest probability has shifted so seriously that the location of point source will generate a large error in no-PSF reconstruction. And azimuthal response only has some slightly blurring not large shift.[19] (c) and (d) are response of center bin and there's no serious blurring.

Using the sinogram blurring, we could get PSF of all voxel. We simulated a sixty-four polygon PET system by GATE software and compared the point response of simulated experimental result, our method and method without blurring matrix. The configuration of the PET system is introduced in section 4. Fig.8 displays two point responses in sinogram (PSF) gained by simulated experiment, our method and ordinary method without blurring matrix. Fig.8 (a), (b) and (c) show the point response of center point. And the point response of simulated experiment, our method and the ordinary method are similar because of the weak blurring showed in Fig.7 (c) and (d). But at the edge of the FOV in Fig.8 (d), (f) and (e), edge bins (the top of sinogram and the bottom of sinogram) begin to blur or spread in the point response of simulated experiment. The point response using our method can describe this blurring spread. While the point response of ordinary method does not contain the blurring spread.

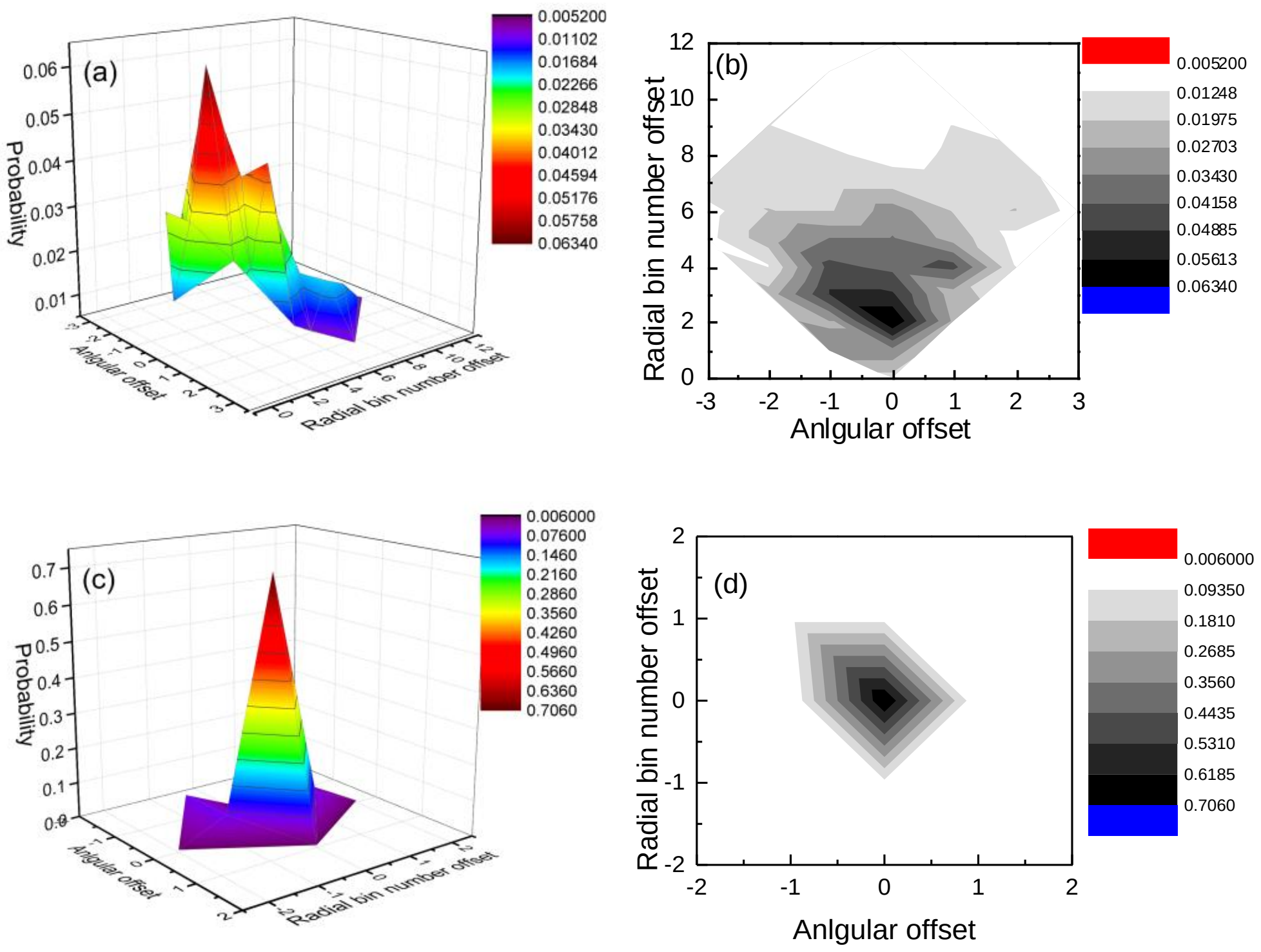

Fig.7. (color online).The blurring responses of LOR at two different position in 0 °sinogram. We take 353 radial bins in each angle for scanner. (a) and (b) represent the NO.1 bin. (c) and (d) represent the NO.176 bin (the center bin). (a) and (c) present the 3D view of the response. (b) and (d) present the plane view of the response.

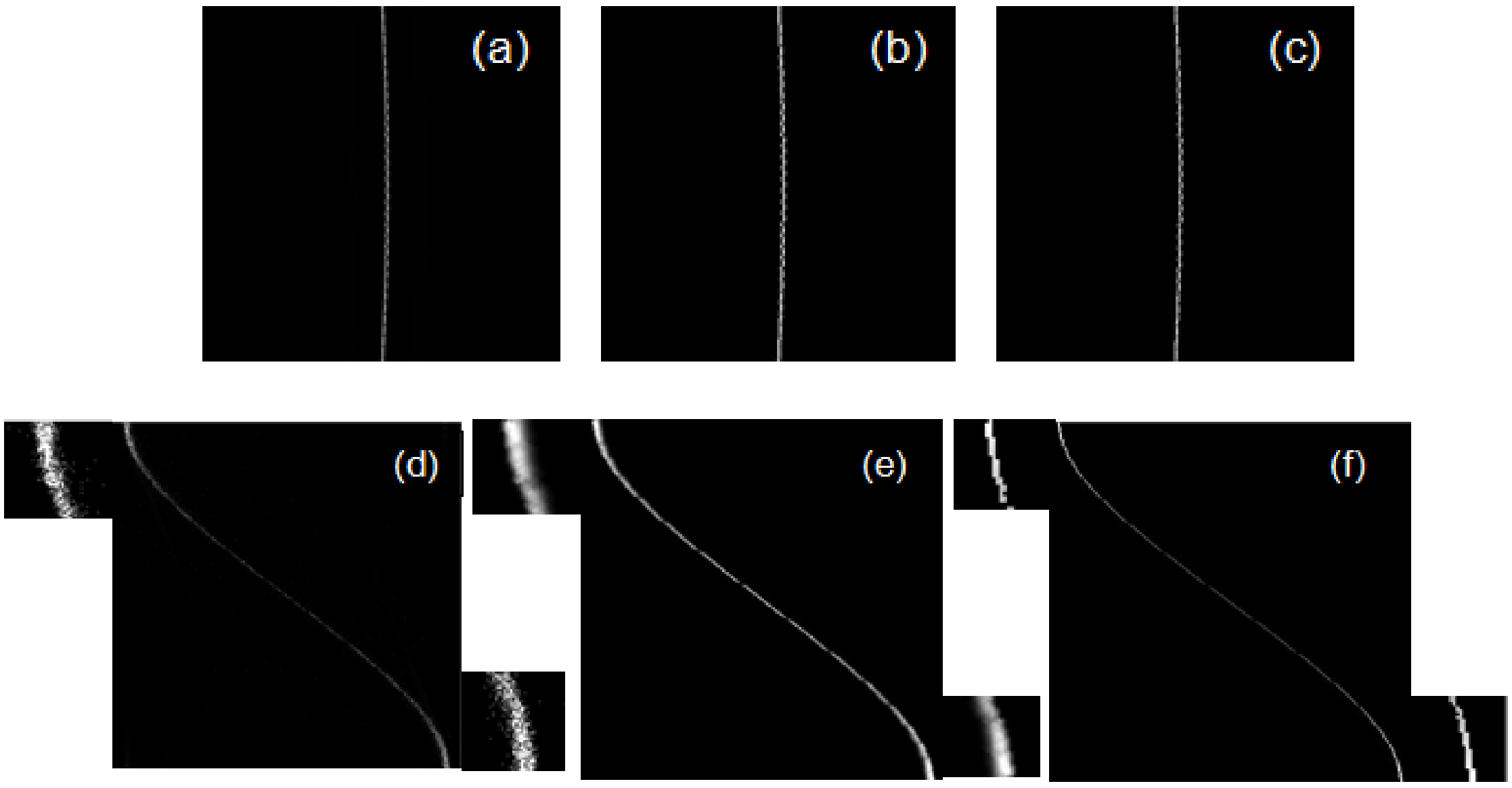

Fig.8. The point responses of two points for three methods in sinogram. Diameters of points are 0.5 mm. (a) and (d) are response of simulated experiment by GATE software. (b) and (e) are results using our method. (c) and (f) are the results of ordinary method without blurring matrix. Top row shows the point response of the point placed in the center of the FOV. Bottom row shows the point response of the point placed in the 300 mm off the FOV center.

### 3.3. Reconstruction of PSF-OSEM

At last, we added PSF factor into the ordinary geometrical system matrix for both forward projection and back projection by real-time computation to reduce the memory consumption. We used the ordered subset expectation maximization (OSEM) iterative reconstruction [3] to get the reconstructed image. In addition, PSF-OSEM methods consume more computational time compared to ordinary OSEM method for the additive spread.

## 4. Results

The raw data was acquired from both experiments in the system of MC simulation using GATE software and in our whole body PET imaging experiments (supported by in-beam whole body PET, Institute of High Energy Physics Chinese Academy of Sciences). The scanner systems were both sixty-four polygon with 64×4 blocks (4 blocks in axial direction). Each block equipped with 11×11 LYSO crystals whose size was 3.5 mm×3.5 mm×15 mm as we mentioned above. The gap between every two blocks was 4 mm. The raw data was acquired with a 361 keV–661 keV energy window and a 6 ns timing window. We binned the emission data to a 704×353×87 sinogram matrix after Fourier rebinning[33]. There were 704 angles for every 87 slices and 353 radial bins for each angle in this sinogram matrix.

### 4.1. Image resolution

Fig.9 shows comparison of the reconstructed image of point array of PSF-OSEM and ordinary OSEM. The raw data was simulated in the system of MC simulation. The diameter of each point is 0.5 mm. The smallest and largest distances of the points away from the FOV center are 140 mm and 300 mm. The distance between every two points is 20 mm in both radial and tangential directions. The image pixel was 1 mm. In both reconstruction algorithms, 8 subsets were used and the reconstructions were stopped after 10 iterations. In Fig.9 (a), OSEM result shows an increased radial resolution loss as the radial distance increasing. While PSF-OSEM result shows a more uniform radial resolution. Fig.9 (b) shows the profiles along the middle row. The curve of points acquired by OSEM method show the degeneration of radial resolution and larger positional shift to FOV center which may result in locating error.

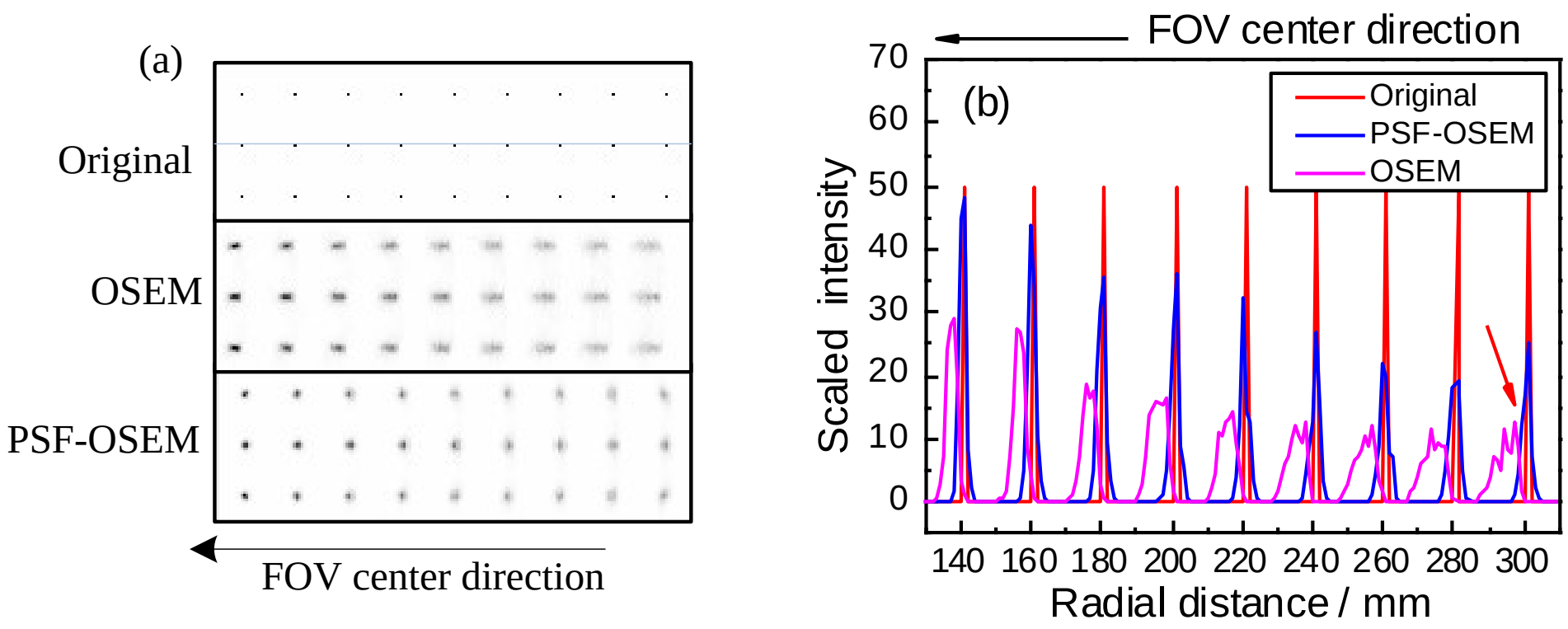

Fig.9. (color online) The reconstructed image and the middle line profile of the point array generated by MC simulation. The closest point is 140 mm off the center of FOV and the furthest point is 300 mm off the center of FOV. The distance between every two points is 20 mm in both radial and tangential directions. (a) is the reconstructed image of point array. (b) is the profile of middle line points.

Fig.10 (a), (b) and (c) shows the curves of resolution and radial position error versus radial distance. The points were generated singly from the system of MC simulation according to section 3 of NEMA Standards Publication.[4] In both reconstruction algorithms, 8 subsets were used and the reconstructions were stopped after 2 iterations. The pixel size was 0.5 mm for the reconstruction. Resolution is specified as the full width at half maximum (FWHM) of the point source response.[4] In Fig.10 (a), the best and worst radial resolutions for OSEM results are 2.46 mm and 6.96 mm. While for PSF-OSEM, the best radial resolution is 2.42 mm and the worst one is 4.41mm. The radial resolution has been improved for the PSF-OSEM reconstruction. Fig.10 (b) shows the similar tangential resolutions in the two methods. But tangential resolution of point at 300 mm radial distance suddenly becomes lower for the OSEM reconstruction. This is because the furthest point has turned long and narrow even split into two or three points because of DOI effects. In Fig.9 (b), the profile curve of the point on 300 mm radial distance shows the split and Fig.10 (d) shows the narrow tangential resolution. Fig.10 (c) shows the larger error in radial position for OSEM reconstruction. Most of the points shift towards the FOV center and the biggest shift is 6mm. In Fig.10 (c), the curve also shows saltus at 150mm radial distance and 300 mm radial distance in OSEM algorithm result as well as the sudden slight decrease of tangential resolution in Fig.10 (b). That is because both two positions are at block gap location where the DOI effects are serious. These saltus disappear in the result of PSF-OSEM in comparison. Table 3 shows the percent of image resolution improved. We define the percent as follow:

$$I = \frac{R_{OSEM} - R_{PSF}}{R_{OSEM}} \times 100\% . \qquad (3)$$

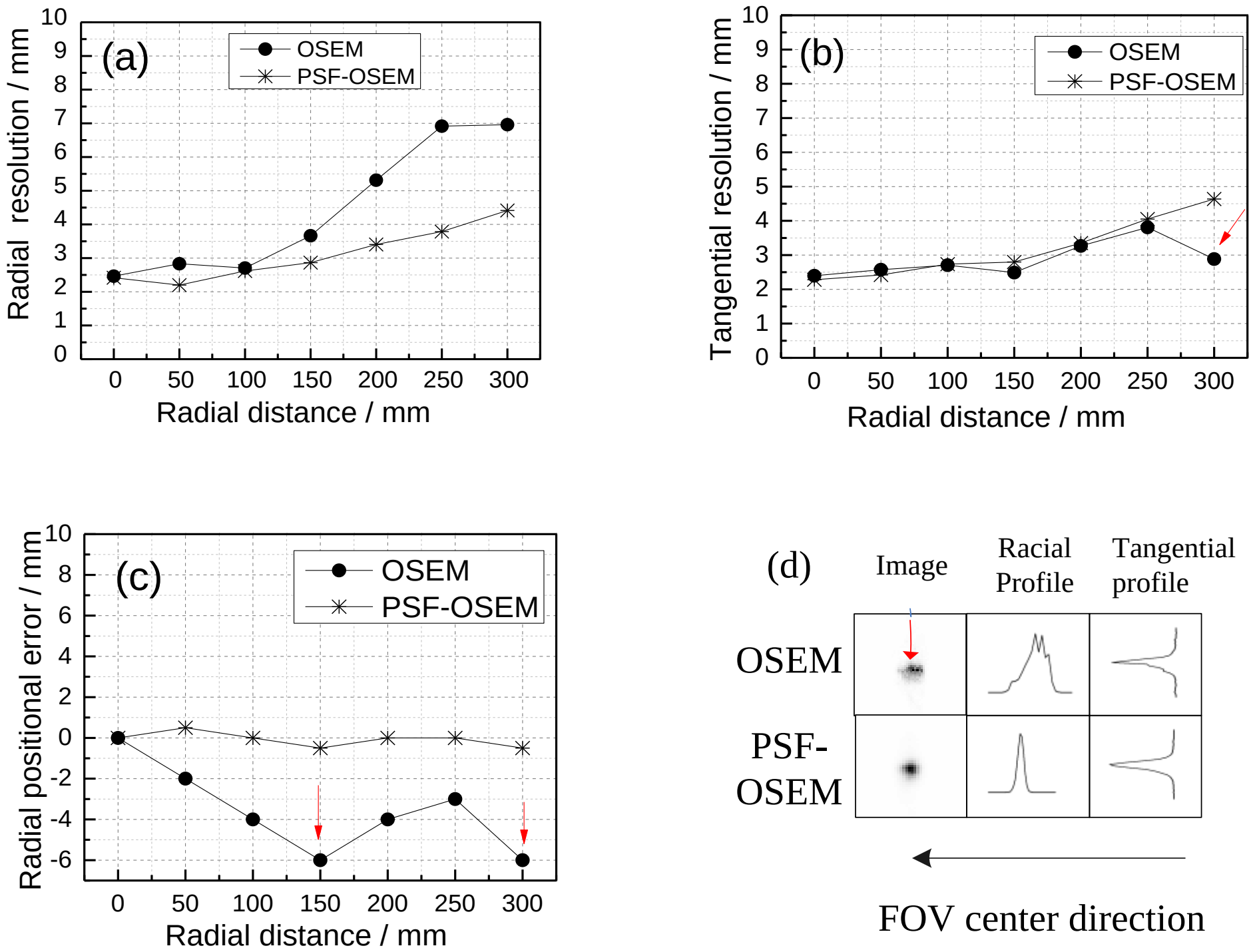

Fig.10. (color online) The curves of radial resolution, tangential resolution and the radial positional error versus radial distance of single points generated from the system of MC

simulation. (a) is the radial resolution curve. (b) is the tangential resolution curve. (c) is the radial positional error curve. (d) is the reconstructed image of the 300 mm point(the furthest point).

Table 3. The percent of image resolution improved for simulated data

| Improved item | Radial offset (mm) | | | | | | |
|---|---|---|---|---|---|---|---|
| | 0 | 50 | 100 | 150 | 200 | 250 | 300 |
| Radial resolution | 1.9% | 22.35% | 3.2% | 21.8% | 35.9% | 45.2% | 36.6% |
| Tangential resolution | 4.8% | 5.9% | -1.1% | -12.4% | -2.5% | -6.4% | -60.8% |

Figs.11 and 12 show resolution condition of our PET data. In Fig.11, the activity concentration of the top $^{68}Ge$ rod source was 3 μCi and the other two $^{68}Ge$ rob sources were 0.5 μCi. The diameter of each rod was 3 mm. The center of triangle composed of three robs was at the position of 282 mm away from the center of the FOV. We made 8 subsets and 10 iterations and 1 mm image pixel for both two reconstructions. Fig.12 (a), (b) and (c) are curves of image resolution and radial positional error versus radial distance of the single points respectively. The diameter of 25 μCi $^{22}Na$ single point was 0.5 mm and the position was configured according to section 3 of NEMA Standards Publication.[4] We took 8 subsets and 2 iterations without smoothing both in the two algorithms. The pixel size was also 0.5 mm according to NEMA Standards Publication. [4] We can come to the same conclusion in experimental data and simulated data. The smallest and biggest radial resolution are 2.39 mm and 4.24 mm respectively in PSF-OSEM result. And the corresponding radial resolutions are 2.98 mm and 7.14 mm respectively in OSEM result. Table 4 shows the percent of the image resolution improved for our PET dada.

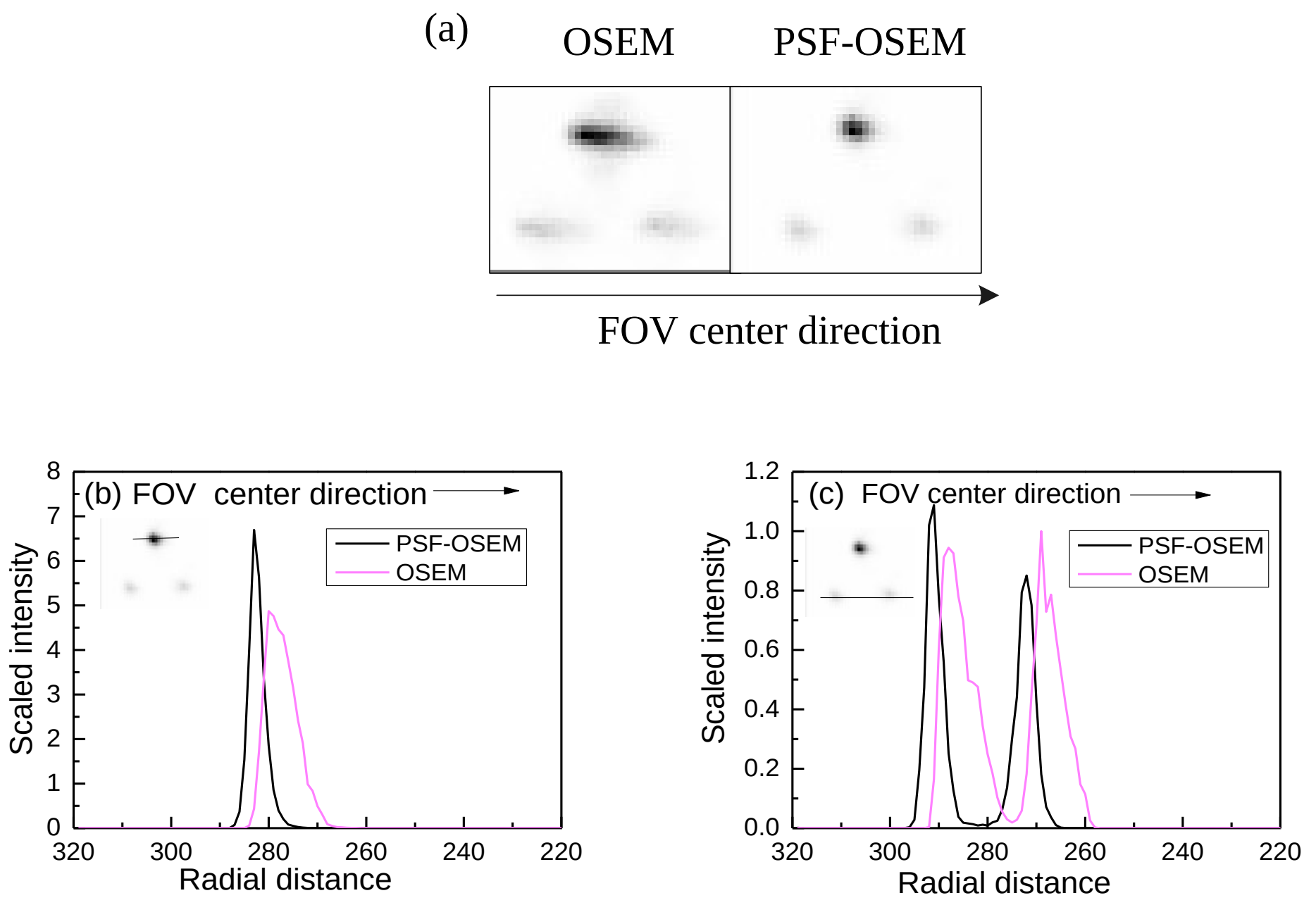

Fig.11. (color online) The center slice of the reconstructed image of the three rob source data acquired from our PET scanner. The center of triangle composed of the three robs is at the position of 282 mm away from the center of the FOV. (a) is the center slice of reconstructed image. (b) is the profile of the top rob of the center slice. (c) is the profile of the other two robs of center slice.

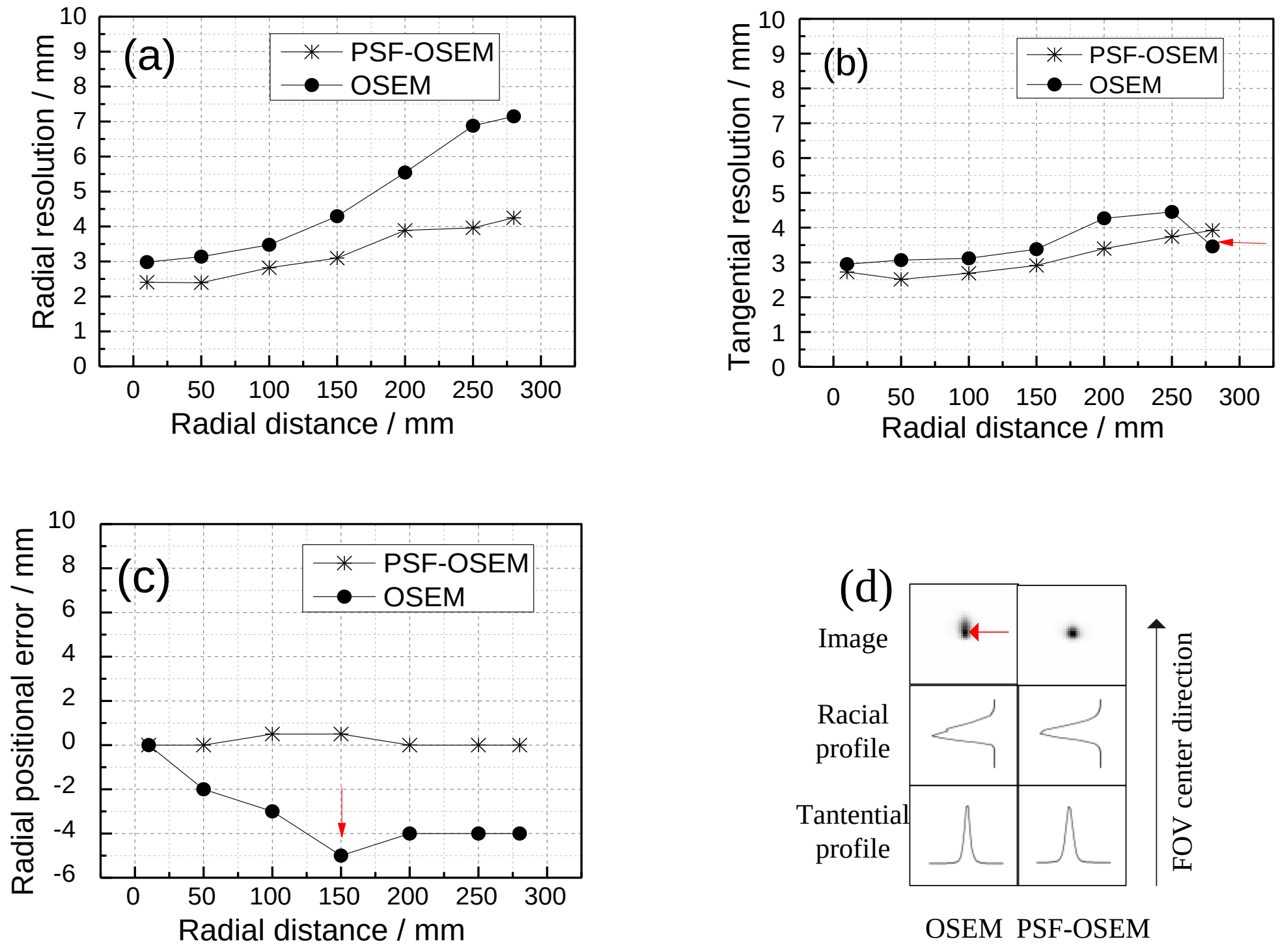

Fig.12. (color online) The curves of radial resolution, tangential resolution and radial positional error versus radial distance of the single points data acquired from our PET scanner.. (a) is the radial resolution curve. (b) is the tangential resolution curve. (c) is the radial positional error curve. (d) is the reconstructed image of the 280 mm point(the furthest point).

Table 4. The percent of image resolution improved for our PET data

| Improved item | Radial offset(mm) | | | | | | |
|---|---|---|---|---|---|---|---|
| | 10 | 50 | 100 | 150 | 200 | 250 | 280 |
| Radial resolution | 19.3% | 23.8% | 18.8% | 27.9% | 29.8% | 42.4% | 40.6% |
| Tangential resolution | 7.6% | 20% | 13.8% | 13.8% | 20.4% | 15.9% | -13.4% |

## 4.2. The contrast recovery and noise property

We simulated a sphere phantom in the system of MC simulation. The ratio of activity concentration of hot spheres and background was 8:1. The diameters of hot spheres were 10 mm, 13 mm, 17 mm, 22 mm respectively. Fig.13 shows the center transverse slice of reconstructed image and the center row profile of the two smallest spheres. The raw data was reconstructed by 8 subsets and 5 iterations without smoothing after corrections of scatter and attention. The image pixel size was 1 mm. Fig 14 (a) shows the ROI chosen method. Fig.14 (b) and (c) show the contrast recovery (hot sphere ROI mean divide the background ROI mean) curve of the hot spheres and the percent background variability (background ROI std divide the background ROI mean) curve versus sphere diameter. The percent background variability is usually used to evaluate the noise property.[4] Table 4 shows the percent of hot sphere contrast recovery and percent background variability improved.

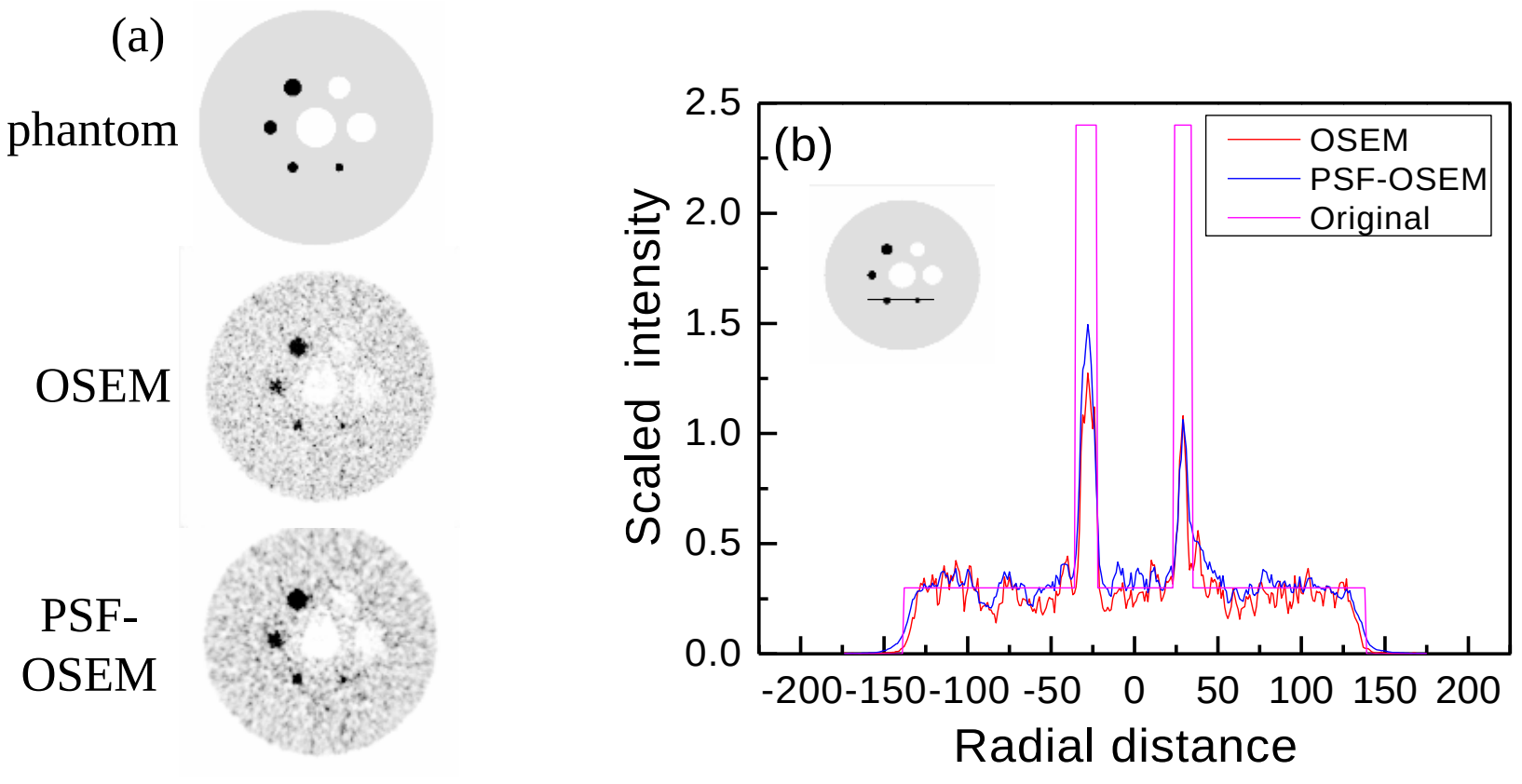

Fig.13. (color online) Transverse view of the center slice and a profile of center row of the two smallest spheres of the sphere phantom, which simulated by the system of MC simulation. The ratio of activity concentration of hot spheres and the background is 8:1. The hot spheres' diameter are 10 mm, 13 mm, 17 mm and 22 mm respectively. (a) is the reconstructed image. (b) is the profile of center row of the two smallest spheres.

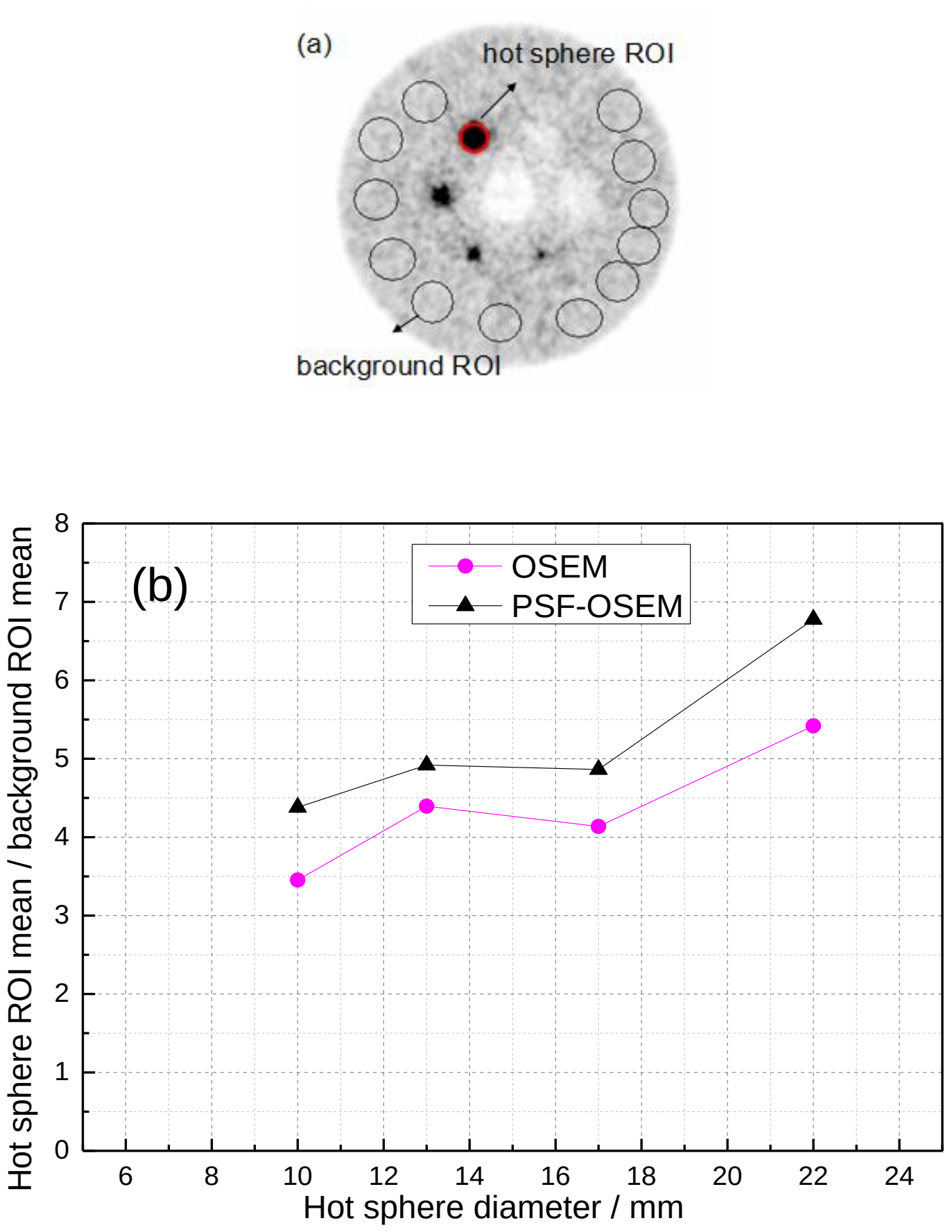

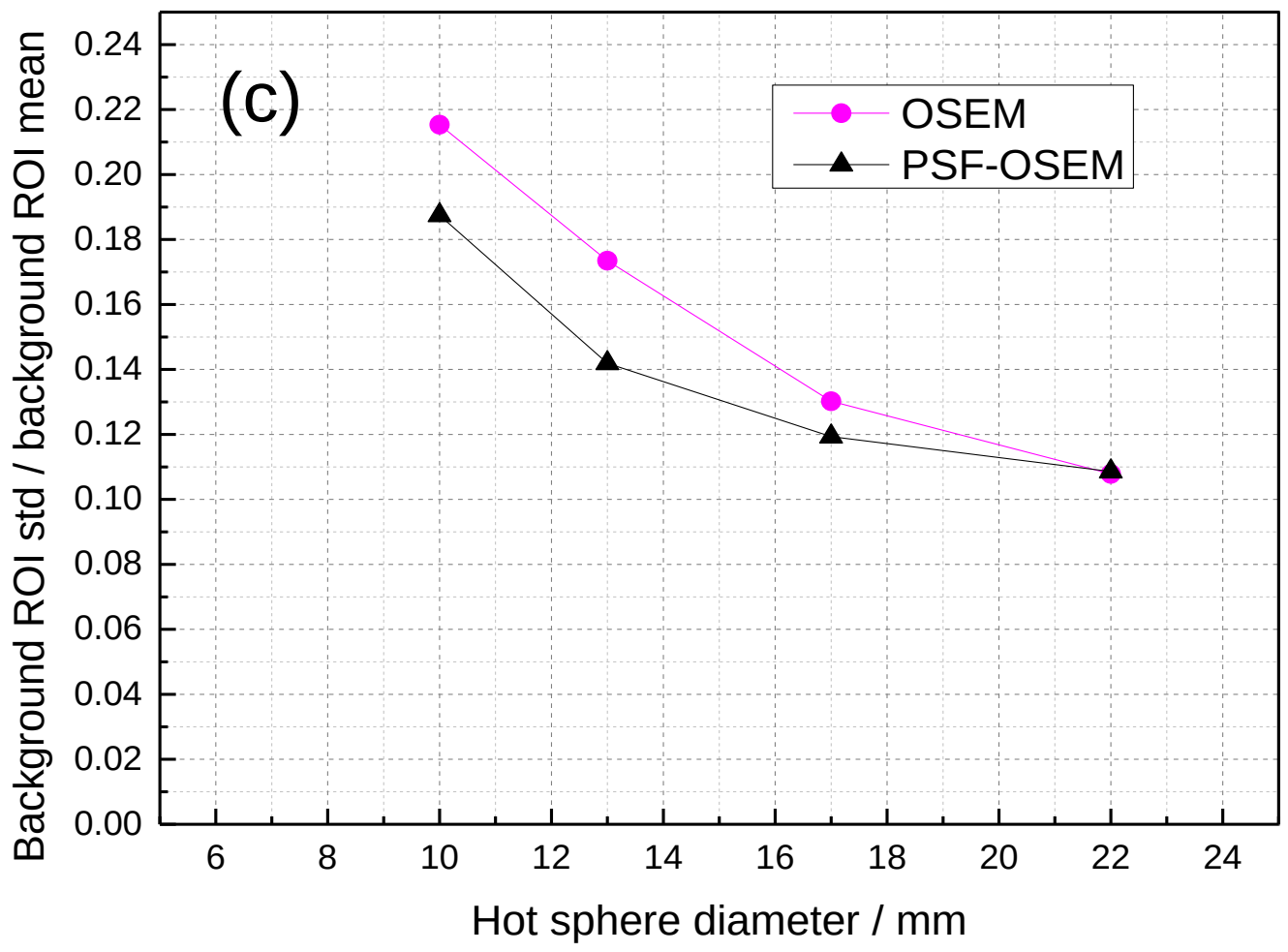

Fig.14. (color online) (a) is the ROI chosen method. (b) is the hot sphere contrast recovery (hot sphere ROI mean divide the background ROI mean) curve versus the sphere diameter. (c) is the percent background variability (background ROI std divide the background ROI mean) versus the sphere diameter.

Table 5. The percent of hot sphere contrast recovery and percent background variability improved

| Improved Item | Sphere diameter(mm) | | | |
|---|---|---|---|---|
| | 10 | 13 | 17 | 22 |
| Hot sphere contrast recovery | 26% | 11.9% | 10.6% | 25% |
| Percent background variability | 12.9% | 18.1% | 8.3% | 0% |

## 5. Conclusions and discussions

In this paper, we proposed a new method of PSF iterative reconstruction. Our method shows good result with improved image radial resolution, contrast recovery and noise property. Moreover, single photon response function in this method depends on configuration of crystal instead of the system geometry. With this advantage, we only need to change the calculation coincidence process to adapt to different PET system geometry. However, there are also several problems need to be discussed.

### 5.1. The single photon incidence angle step

In our paper, we chose 5 ° for the angle step and image radial resolution improved inspiringly. If we resize the step for a more proper value, the result will be better. In this paper, we simulated the response by a uniform step. In fact the response of single photon penetrating crystals may be nonuniform. The distribution of single photon penetrating crystals will be studied in future.

### 5.2. Convergence and computational time

Commonly, PSF reconstruction converges slower than the non-PSF because PSF contains a lot blurring information. We must consider to add some accelerated algorithms (for example accelerated algorithm based on GPU) in PSF reconstruction to solve this problem.

### 5.3. Influence of the nonuniformity of reality detector unit

The method has assumed that the crystals are uniform. Actually, the cutting technology is relative maturity， the error among the sizes of the crystals is ±0.05 mm, Our crystal size in this article is 3.5 mm×3.5 mm, the error is less than 1.4%. And we take the tube model method to make the system matrix, so this deviation is acceptable.

In addition, we also take standard regular polygon to make the system matrix. In reality, we took rack of regular polygon to assemble the detector structure. The machining error of the rack of regular polygon is less than 0.1 mm, and the assembly error is less than 0.5 mm. If we consider all these error, the total error of the detector structure is less than 1 mm. In the PET system of our particle, the size of each block is 39.7 mm. So each crystal is in the right place as in our model within the margin of error.

## References

[1] Gambhir S S, Czermin J, schwimmer J, Silverman D H, Coleman R E and Phelps M E 2001 *J. Nucl. Med* 42 1S-93S

[2] Natterer F and Wuebbeling F 2001 *Mathematical methods in image reconstruction* (Vol. 5) (Philadelphia, PA: SIAM) p.81

.[3] Hudson H and Larkin R 1994 *IEEE Trans. Med. Imaging* 13 601

[4] Daube-Witherspoon M E, Karp J S and Casey M E 2002 *J. Nucl. Med.* 43 1398

[5] Mawlawi O and Townsend D W 2009 *Eur J. Nucl. Med. Mol. Imaging* 36 S15

[6] Nestle U, Weber W, Hentschel M and Grosu A L 2009 *Phys. Med. Biol.* 54 R1

[7] Pan T and Mawlawi O 2008 *Med. Phys.* 35 4955

[8] Wiant D B, Gersh J A, Bennett M C and Bourland J D 2009 *Nuclear Science Symposium Conference Record*, October 24 - November 1, 2009, Orlando, FL, p.3758

[9] Chien M K, Yun D, Qing G X and Chin T C 2008 *IEEE Trans. Med. Imaging* 1346-1358

[10] Lecomte R, Schmitt D and Lamoureux G, 1984 *IEEE Trans. Nucl. Sci.* NS-31 556

[11] Liang Z 1994 *IEEE Trans. Med. Imaging* 13 314

[12] Rahmim A, Tang J, Lodge M A, Lashkari S, Ay M R, Lautamaki R, Tsui B M W and Bengel F M 2008 *Phys. Med. Biol.* 53 5947

[13] Alessio A M, Kinahan P E and Lewellen T K 2006 *IEEE Trans. Med. Imaging* 25 828837

[14] Mumcuoglu E U, Leahy R M, Cherry S R and Hoffman E 1996 *Nuclear Science Symposium, 1996. Conference Record, 1996 IEEE*, November 2-9, 1996, Anaheim, CA, 3 p.1569

[15] Qi J, Leahy R M, Cherry S R, Chatziioannou A and Farquhar T H 1998 *Phys. Med. Biol.* 43 1001

[16] Alessio A M and Kinahan P E 2008 *5th IEEE International symposium on biomedical imaging: From nano to macro*, May 14-17, 2008, Paris, p.1315

[17] Bernardi E D, Zito F and Baselli G 2007 *Engineering in Medicine and Biology Society, 2007. EMBS 2007. 29th Annual International Conference of the IEEE*, August 22-26, 2007 Lyon, p. 6547

[18] Bernardi E D, Zito F, Michelutti L, Mainardi L, Gerundini P and Baselli G *Engineering in Medicine and Biology Society, 2003. EMBS 2003.25th Annual International*

*Conference of the IEEE*, September 17-21, 2003 1, p. 975.

[19] Fin L, Bailly P, Daouk J and Meye M E 2009 *Med. Phys.* 36 3072

[20] Panin V Y, Kehren F, Michel C and Casey M 2006 *IEEE Trans. Med. Imaging* 25 907

[21] Panin V Y, Kehren F, Rothfuss H, Hu D, Michel C and Casey M E 2006 *IEEE Trans. Nucl. Sci.* 53 152

[22] Tohme M S and Qi J 2009 *Phys. Med. Biol.* 4 3709

[23] Alessio A M, Stearns C W, Tong S, Ross S G, Ganin A and Kinahan P E 2010 *IEEE Trans. Med. Imaging* 29 938

[24] Qi J, Leahy R, Cherry S R, Chatziioannou A and Farquh-ar T 1998 *Phys. Med. Bio.* 43 1001

[25] Wang L, Wu L W, Wei L, Gao J, Sun C L Chai P and Li D W 2014 *Chin. Phys. B* 23 2027802

[26] Badawi R D, Lodge M and Marsden P K 1998 *Phys. Med. Bio.* 43 189

[27] Tohme M S 2011 Iterative Image Reconstruction for Positron Emission Tomography Based on Measured Detector Response Function Tomography Based on Measured Detector Response Function (Ph.D. dissertation) (California :University of California )

[28] Joseph P M 1982 *IEEE Trans. Med. Imaging* MI-1 192

[29] Johnson C, Yan Y, Carson R, Martino R and Daube-Witherspoon M 1995 *IEEE Trans. Nucl. Sci.* 42 1223

[30] Chen C, Lee S and Cho Z 1991 *IEEE Trans. Med. Imaging* 10 513

[32] Leonard S 2005 Spatial resolution Study of PET Detector Modules Based on LSO Crystals and Avalanche Photodiode Arrays (Ph.D. dissertation) (Brussel:Vrije Universiteit Brussel)

[31] http://www.opengatecollaboration.org/

[33] Defrise M, Kinahan P, Townsend D, Michel C, Sibomana M and Newport D 1997 *IEEE Trans. Med. Imaging* 16 145